\documentclass{article}
\usepackage{arxiv}

\usepackage{graphicx}
\usepackage{epstopdf}
\usepackage{dcolumn}
\usepackage{bm}
\usepackage{subfigure}
\usepackage{amsmath}
\usepackage{epsfig}
\usepackage{verbatim}
\usepackage{float}

\usepackage[utf8]{inputenc} % allow utf-8 input
\usepackage[T1]{fontenc}    % use 8-bit T1 fonts
\usepackage{hyperref}       % hyperlinks
\usepackage{url}            % simple URL typesetting
\usepackage{booktabs}       % professional-quality tables
\usepackage{amsfonts}       % blackboard math symbols
\usepackage{nicefrac}       % compact symbols for 1/2, etc.
\usepackage{microtype}      % microtypography
\usepackage{lipsum}
\usepackage{graphicx}
\graphicspath{ {./images/} }

\title{Effect of heterogeneous risk perception on information diffusion, behavior change, and disease transmission}

\author{
 Yang Ye \\
  City University of Hong Kong\\
  Hong Kong SAR, China\\
   \And
 Qingpeng Zhang \\
  City University of Hong Kong\\
  Hong Kong SAR, China\\
  \And
 Zhongyuan Ruan \\
  Institute of Cyberspace Security\\
  Zhejiang University of Technology\\
  Hangzhou, China \\
  \And
  Zhidong Cao \\
  The State Key Laboratory of Management and Control for Complex Systems\\
  Institute of Automation\\
  Chinese Academy of Sciences\\
  Beijing, China
  \And
  Qi Xuan \\
  Institute of Cyberspace Security\\
  Zhejiang University of Technology\\
  Hangzhou, China\\
  \And
  Daniel Dajun Zeng \\
  The State Key Laboratory of Management and Control for Complex Systems\\
  Institute of Automation\\
  Chinese Academy of Sciences\\
  Beijing, China
}

\begin{document}
\maketitle

\begin{abstract}
Motivated by the importance of individual differences in risk perception and behavior change in people's responses to infectious disease outbreaks (particularly the ongoing COVID-19 pandemic), we propose a \textbf{h}eterogeneous \textbf{D}isease-\textbf{B}ehavior-\textbf{I}nformation  (\textbf{hDBI}) transmission model, in which people's risk of getting infected is influenced by information diffusion, behavior change, and disease transmission. We use both a mean-field approximation and Monte Carlo simulations to analyze the dynamics of the model. Information diffusion influences behavior change by allowing people to be aware of the disease and adopt self-protection, and subsequently affects disease transmission by changing the actual infection rate. Results show that (a) awareness plays a central role in epidemic prevention; (b) a reasonable fraction of ``over-reacting" nodes are needed in epidemic prevention; (c) $R_0$ has different effects on epidemic outbreak for cases with and without asymptomatic infection; (d) social influence on behavior change can remarkably decrease the epidemic outbreak size. This research indicates that the media and opinion leaders should not understate the transmissibility and severity of diseases to ensure that people could become aware of the disease and adopt self-protection to protect themselves and the whole population.

\end{abstract}

\maketitle
\section{Introduction} \label{introduction}

People's responses to infectious diseases could greatly affect the transmission patterns of diseases, and information about the transmissibility and severity of the disease conveyed through the media and opinion leaders plays a central role in raising awareness and influencing people's decision-making on whether or not to adopt self-protection (i.e., taking recommended practices to reduce the risk of infection, such as wearing a mask and washing hands with hanitizer in the COVID-19 context \cite{Senn:2009}) \cite{ferguson2007capturing,Rubinb2651,Lau864,RePEc:uwp:jhriss:v:31:y:1996:i:3:p:611-630,article_AIDS,collinson2014modelling}. Here, opinion leaders refer to individuals whose opinions are widely accepted by other people \cite{van2020using,9043580}.

Many studies \cite{perra2012activity,wang2014multiple,ruan2003dynamical,WANG20151,kabir2020impact,han2020individuals,guo2013epidemic} used mathematical models to investigate how disease awareness affects the outbreaks of diseases. Most approaches \cite{fenichel2011adaptive,article_Media_Reports} explored this problem by modifying the parameters in standard epidemic models. Funk et al. \cite{funk2009spread} first incorporated the effect of awareness into classic epidemic models and found that the spread of awareness could substantially reduce the epidemic outbreak size. Gross et al. \cite{PhysRevLett.96.208701} studied epidemic dynamics on an adaptive network, in which susceptible nodes avoid contact with infected nodes by rewiring their connections. Wu et al. \cite{article} modeled the effect of three forms of awareness: global awareness, local awareness, and contact awareness. They showed that only global awareness cannot decrease the likelihood of an epidemic outbreak. Chen \cite{CHEN2009125} examined how the amount of information affects behavior change, and showed that increasing the amount of information that people possess may lower the likelihood of disease eradication.

Multiplex (also named multilayer) networks \cite{article_Layered_Complex_Networks,Multilayer} have been developed to model the dynamic interactions between the spread of information and infection. The spreading of information and infection is represented by multiple network layers. For example, Granell et al. \cite{PhysRevLett.111.128701} proposed the susceptible-infected-susceptible unaware-aware-unaware (SIS-UAU) model that can capture the critical point of the disease outbreak determined by the topological structure of the virtual information diffusion network. It can be extended to many model variants, such as the multiple-information model \cite{PAN201845} that incorporates more than one type of information; the local awareness controlled contagion spreading model \cite{PhysRevE.91.012822}, in which the awareness transition is further influenced by the extent of the awareness of all neighbors; and the susceptible-infected-recovered unaware-aware (SIR-UA) model \cite{KABIR2019118} that also considers the recovered state. In addition to information diffusion and disease transmission, the transmission of protective behavior has been simulated \cite{mao2014modeling}.

Recent studies examined the effect of the media on infectious disease epidemics \cite{granell2014competing, perra2011towards,liu2007media,wang2014filippov,dubey2015proceedings,CRAMER2016739,song2018global,song2019analysis}. Liu et al. \cite{liu2007media} proposed a mechanism to illustrate the effect of the media by incorporating the reported numbers of infectious and hospitalized individuals into classic epidemic models. Wang and Xiao \cite{wang2014filippov} used a threshold model where the media can exhibit its effect only when the number of reports reaches a certain value. Dubey et al. \cite{dubey2015proceedings,CRAMER2016739} discussed the optimal amount of information that can not only suppress the transmission of disease but also prevent ``media-fatigue." Song and Xiao \cite{song2018global,song2019analysis} further considered the delay of media effects on people's responses.

To summarize, existing studies indicate that the interplay between awareness and social network structure could greatly influence the transmission of infectious diseases. However, few studies have considered individual differences in people's responses to messages conveyed through the media and opinion leaders during epidemics. People's responses are not only influenced by the transmissibility and severity of the disease, but also the personal risk perception \cite{sjoberg2000factors,wahlberg2000risk,RENNER2015702}. Here, risk perception refers to the subjective judgment about the risk of the disease. Similar to smart nodes in information diffusion \cite{ruan2018information,ruan2012epidemic}, individuals who are more fearful of being infected actively engage in self-protection and information sharing. We label these people as ``over-reacting" as compared with ``under-reacting" people who have a relatively low risk perception. This classification can also be applied to the novel coronavirus (COVID-19) epidemic, in which a clear disparity in risk perception causes the diverse reactions of individuals regarding control measures \cite{zhu2020novel,wang2020novel,ASMUNDSON2020102196,kai2020universal}. Lessons from the COVID-19 outbreak in many countries imply that people responding properly to the disease is more important than simple awareness.

Motivated by the importance of individual differences in risk perception and behavior change in people’s responses to infectious disease outbreaks (particularly the ongoing COVID-19 pandemic), we propose a \textbf{h}eterogeneous \textbf{D}isease-\textbf{B}ehavior-\textbf{I}nformation (\textbf {hDBI}) transmission model, which consists of information diffusion, behavior change, and disease transmission. This model aims to describe how different types of nodes (``over-reacting" versus ``under-reacting") influence the prevalence of protective behavior and the epidemic outbreak. The contribution of this study is threefold: First, this is the first quantitative study that examines the incorporation of risk perception and disease-behavior-information transmission on a multilayer network. Second, we consider individual differences in the responses to disease-related information originating from variations in the distribution of risk perception among people. Third, we study analytically and numerically the effect of such differences on the epidemic. 

The rest of the paper is organized as follows. First, we describe the model details in section \ref{model}. Second, we adopt the mean-field method \cite{sahneh2013generalized} to formulate the problem mathematically in section \ref{Theoretical analysis}. Third, we perform extensive experiments with Poisson degree distribution-based Erd\H{o}s–R\'enyi networks and explore the effects of different parameters on the epidemic outbreak in section \ref{Results}. Lastly, we conclude the paper with discussions of future work in section \ref{Conclusion}.

\section{Model} \label{model}
\begin{figure}
\includegraphics[width=1.0\linewidth]{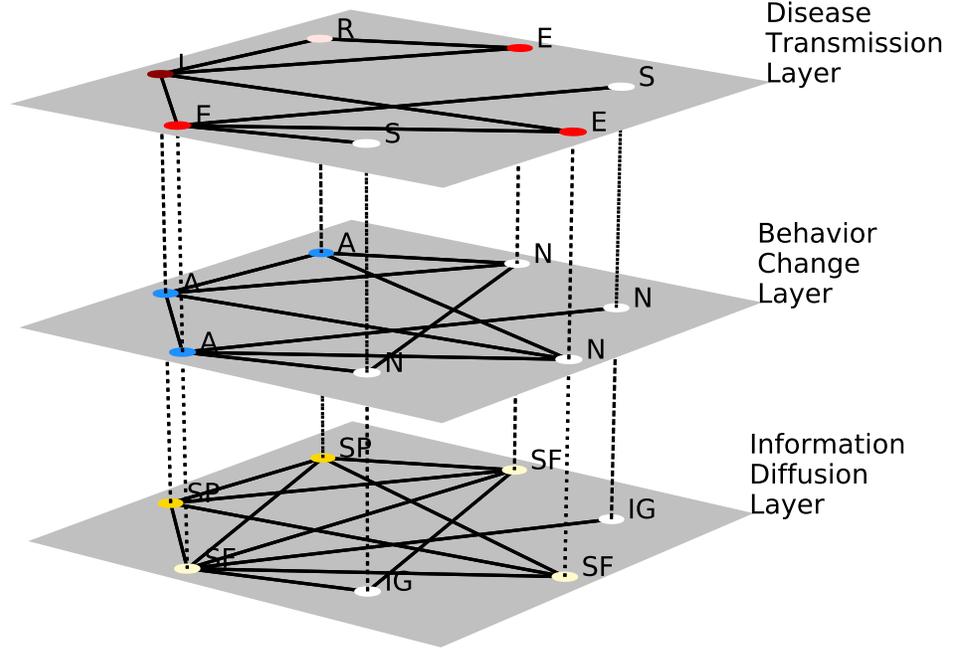}
\caption{\label{fig:3_layer_model} Structure of the three-layer network framework. The bottom, middle, and top layers represent information diffusion, behavior change, and disease transmission, respectively.}
\end{figure}

We propose a three-layer network model, namely, the \textbf{h}eterogeneous \textbf{D}isease-\textbf{B}ehavior-\textbf{I}nformation (\textbf {hDBI}) transmission model, to incorporate information diffusion, behavior change, and disease transmission. Here, nodes may become aware of the risk of getting infected and then change their behavior by adopting self-protection, which will further affect the disease transmission. We first introduce the details as shown in Fig.~\ref{fig:3_layer_model}. All layers in the model contain the same set of nodes.   

\begin{figure}
\includegraphics[width=1.0\linewidth]{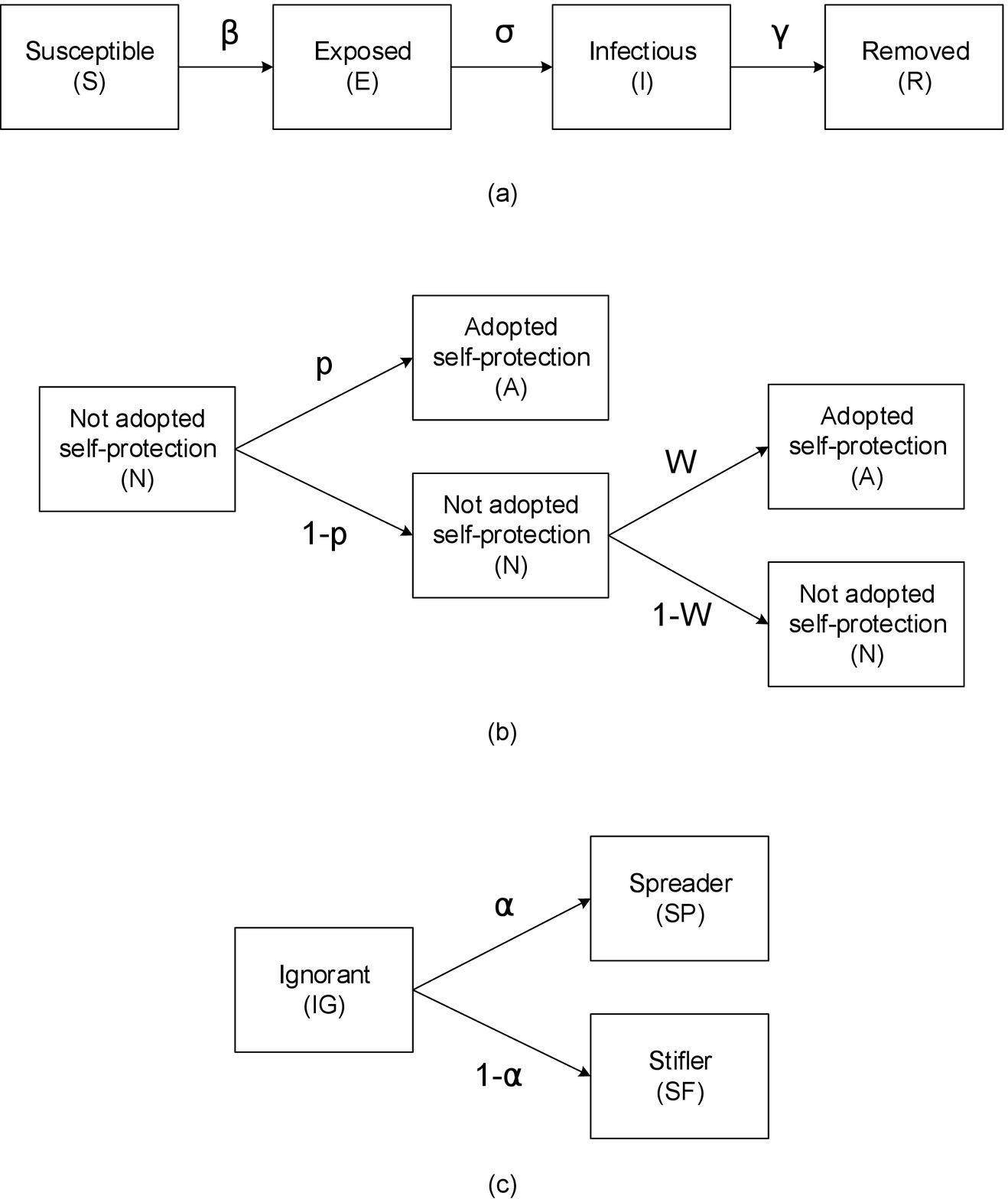}
\caption{\label{fig:SEIRandSIR} Illustration of the transitions between states on (a) the \textit{Disease} \textit{Transmission} \textit{Layer}, (b) the \textit{Behavior} \textit{Change} \textit{Layer}, and (c) the \textit{Information} \textit{Diffusion} \textit{Layer}. The $Ignorants$ ($IG$) are unaware of the disease and the $Spreaders$ ($SP$) and $Stiflers$ ($SF$) have become aware of the disease. $\alpha$ and $p$ correspond to $\alpha_o$ and $p_o$ for ``over-reacting" nodes, and $\alpha_u$ and $p_u$ for ``under-reacting" nodes, respectively.}
\end{figure}

On the bottom layer, namely, the \textit{Information Diffusion Layer},  edges represent the social contacts through which nodes can share the disease-related information. The sources of the information are the media and opinion leaders. Nodes are divided into three classes on this layer: $Ignorants$, $Spreaders$, and $Stiflers$ denoted by $IG$, $SP$, and $SF$, respectively. $Ignorants$ are those who are \textit{unaware} of the disease-related information and can become $Spreaders$ (with a probability $\alpha$) or $Stiflers$ (with a probability $1-\alpha$) if at least one neighbor is a $Spreader$. $Spreaders$ and $Stiflers$ are those who are \textit{aware} of the disease. We assume that the states of $Spreaders$ and $Stiflers$ do not change for simplification. Once a $Spreader$, the node will keep spreading the information to all its neighbors in each period. $Stiflers$, on the other hand, do not further spread the information. We present the scheme mentioned above in panel (c) of Fig.~\ref{fig:SEIRandSIR}. 

Each piece of information has an alarming level $y$, which represents the transmissibility and severity of the disease conveyed through the media and opinion leaders. Each node on the network has a personal risk perception, a constant parameter $x_i$ for node $i$. By comparing the values of $y$ and $x_i$, we classify nodes into two sets: $\{i\mid y \ge x_i\}$, ``over-reacting", and $\{i\mid y < x_i\}$, ``under-reacting". ``Over-reacting" nodes have a higher probability to spread the information as follows
\begin{equation}
\alpha=\left\{
\begin{aligned}
\alpha_o \quad y\ge x_i\\
\alpha_u \quad y<x_i
\end{aligned}
\right.
\label{inforate}
\end{equation}where $0\leq\alpha_u <\alpha_o\leq1$. 

On the middle layer, namely, the \textit{Behavior Change Layer}, edges represent the social contacts through which a node can observe other nodes' states of behavior change. Nodes on this layer have two states: adopted self-protection ($A$) and not adopted self-protection ($N$). $Ignorants$ on the \textit{Information Diffusion Layer} are always in the $N$ state on the \textit{Behavior Change Layer} because they are unaware of the risk. $Spreaders$ and $Stiflers$ have a tendency $p$ to change the behavior by adopting self-protection, and ``over-reacting" nodes have higher behavior change tendency than ``under-reacting" nodes as follows
\begin{equation}
p=\left\{
\begin{aligned}
p_o \quad y\ge x_i\\
p_u \quad y<x_i
\end{aligned}
\right.
\label{protectrate}
\end{equation}
where $0\leq p_u <p_o\leq1$. 

The probability of a node to change behavior is not only dependent on $p$, but also the behavior of its neighbors. The behavior change result can be determined in two steps. First, we use the tendency $p$ to determine if the node changes the behavior independent from the social influence. If the node does not change the behavior, then the model further checks its neighbors' states. If more than half of the nodes in its ego network (the node and all its neighbors) have changed behavior, then the node will be influenced to change as well. Let $k$ denote the degree of a node and $z$ denote the number of its neighbors who have already changed behavior, we use $\varphi$ to summarize all neighbors' states as follows

\begin{equation}
\varphi=\left\{
\begin{aligned}
1\quad z > k-z+1\\
0\quad z \leq k-z+1,
\end{aligned}
\right.
\end{equation} the smallest integer value of $z$ satisfying $\varphi=1$ is $\lceil{\frac{k+2}{2}}\rceil$, thus the overall behavior change probability is as follows
\begin{equation} \mathcal{P}=p+(1-p)W,
\label{behavior change probability}\end{equation}where $W=P(\varphi=1)=P(z \geq \lceil{\frac{k+2}{2}}\rceil)$. We present the scheme mentioned above in panel (b) of Fig.~\ref{fig:SEIRandSIR}.

The top layer, \textit{Disease Transmission Layer}, is modeled by the generalized Susceptible-Exposed-Infectious-Recovered (SEIR) model. Edges represent the physical contacts through which diseases may transmit. The susceptible, exposed, infectious, and recovered nodes are denoted by $S$, $E$, $I$, and $R$, respectively. A susceptible node can be infected by either an infectious node with an infection rate $\beta_I$ or an exposed node with an infection rate $\beta_E$. In the classic SEIR model, exposed nodes are infected but do not develop symptoms nor become infectious. Here, we allow the transmission from exposed nodes to model the widely reported asymptomatically-infected cases in the COVID-19 pandemic and other similar disease outbreaks \cite{wilder2005asymptomatic,NEJMc2001468,papenburg2010household}. If we set $\beta_E=0$, then this layer is identical to the classic SEIR model. For simplicity and consistency with previous work \cite{hou2020effectiveness,wu2020nowcasting}, the hDBI model ignores the transition from asymptomatically-infected to recovered. However, the model is flexible in incorporating various asymptomatic scenarios.

The infection rate $\beta$ (including $\beta_I$ and $\beta_E$) is a constant value only related to the disease. The actual infection rate could vary because of different self-protection outcomes. Considering an edge connecting two nodes, we use $n \in \{0,1,2\}$ to denote the number of nodes who have adopted self-protection. The actual transition rate is $\eta^n\beta$, where $\eta$ is the ineffectiveness of self-protection behavior. $\eta \in [0,1]$. $\eta$ is $0$ if the self-protection behavior is fully protective, and $1$ if the self-protection behavior is entirely useless. An exposed node has a transition rate $\sigma$ to become infectious, so the incubation period is $1/\sigma$.  An infectious node has a transition rate $\gamma$ to become recovered and immune to the same disease. All possible state transitions are shown in panel (a) of Fig.~\ref{fig:SEIRandSIR}.

\section{Theoretical analysis} \label{Theoretical analysis}
We adopt the mean-field approximation approach to analyze the dynamics of the hDBI model. This technique assumes independence among the random variables and helps us study the complex stochastic model with a simpler model \cite{6423227,barabasi1999mean,li2012susceptible}. For simplicity, we consider a three-layer network with the same topological structure. We consider an Erd\H{o}s–R\'enyi network of $N$ nodes with a Poisson degree distribution $P(k)$, where $k$ denotes the degree of a node. 

On the \textit{Information Diffusion Layer}, $\rho_{IG}(t)$, $\rho_{SP}(t)$, and $\rho_{SF}(t)$ denote the fraction of $Ignorants$, $Spreaders$, and $Stiflers$, at time $t$, respectively. Thus, $\rho_{IG}(t)+\rho_{SP}(t)+\rho_{SF}(t)=1$. An $Ignorant$ remains unaware only if none of its neighbors are $Spreaders$. Hence, the probability that an $IG$ node with degree $k$ not being informed by any neighbor is $(1-\rho_{SP}(t))^{k}$. Thus, the transition rate for a randomly selected $IG$ node being informed by neighbors is 
\begin{equation}
h(t) = 1-\sum_{k=0}^{k_{max}}P(k)(1-\rho_{SP}(t))^{k},
\label{ht}
\end{equation} where $k_{max}$ denotes the maximum degree of the network.

 Assuming a standard normal distribution of the risk perception $x_i$, we can calculate the fraction of ``over-reacting" nodes by evaluating the corresponding cumulative distribution function at the alarming level $y$, denoted by $a$. Thus, those who are aware of the disease will become $Spreaders$ with a probability $a\alpha_o+(1-a)\alpha_u$ or become $Stiflers$ with a probability $a(1-\alpha_o)+(1-a)(1-\alpha_u)$.
 
On the \textit{Behavior Change Layer}, let $\rho_A(t)$ denote the fraction of nodes having adopted self-protection at time $t$, then the probability for a node with degree $k$ choosing to protect itself because of the social influence from neighbors is
$$ W_k=\left\{
\begin{aligned}
0& & k=0,1 \\
\sum_{b=\lceil{\frac{k+2}{2}}\rceil}^{k}\binom{k}{b}&[\rho_{A}(t)]^{b}[1-\rho_{A}(t)]^{k-b} & otherwise,
\end{aligned}
\right.
$$
which is equal to the probability that at least $\lceil{\frac{k+2}{2}}\rceil$ neighbors have adopted self-protection, thus the probability of changing behavior due to the social influence from neighbors for a randomly selected node is $W=\sum_{k=0}^{k_{max}}P(k)W_k$. Therefore, the probability of changing behavior is $p_o+(1-p_o)W$ for ``over-reacting" nodes, and $p_u+(1-p_u)W$ for ``under-reacting" nodes [according to Eq.~(\ref{behavior change probability})]. Besides, $Ignorants$ will never adopt self-protection, thus, the fraction of nodes who might adopt behavior change at time $t+1$ is $\rho_{SP}(t+1)+\rho_{SF}(t+1)$. So we can get the fraction of nodes having adopted self-protection at time $t+1$ as follows:
 \begin{equation}
 \begin{aligned}
 \rho_A(t+1) =& [\rho_{SP}(t+1)+\rho_{SF}(t+1)]\left.\{ap_o + (1-a)p_u\right.\\
 &\left.+ [a(1-p_o)+(1-a)(1-p_u)]W\right.\},\\
 \end{aligned}
 \label{behavior change}
 \end{equation}
Therefore, 
 \begin{equation}
 \rho_N(t+1) = 1 - \rho_A(t+1),
 \label{N}
 \end{equation}which is the fraction of nodes having not adopted self-protection at time $t+1$.
 
On the \textit{Disease Transmission Layer}, the fraction of nodes in one of the four health states at time $t$ is denoted by $\rho_S(t)$, $\rho_E(t)$, $\rho_I(t)$, and $\rho_R(t)$. $\rho_S(t)+\rho_E(t)+\rho_I(t)+\rho_R(t) =1$. For a single edge between a susceptible node and an infected one, the probability that both have adopted self-protection is $[\rho_A(t)]^2$, and the corresponding actual transition rate is $\eta^2\beta$. The probability that both nodes have not adopted self-protection is $[1-\rho_A(t)]^2$ and the corresponding actual transition rate is $\eta^0\beta$. The probability that only one of them has adopted self-protection is ${1-[\rho_A(t)]^2-[1-\rho_A(t)]^2}$ and the corresponding actual transition rate is $\eta^1\beta$. Thus, the actual infection rate $l(\beta)$ is
\begin{equation}
  \begin{aligned}
  l(\beta)=&[\rho_A(t)]^2\eta^2\beta + [1-\rho_A(t)]^2\eta^0\beta+ \left\{1-[\rho_A(t)]^2\right.\\
  &\left.-[1-\rho_A(t)]^2\right\}\eta^1\beta= \beta[1+(\eta-1)\rho_A(t)]^2.
  \end{aligned}
  \label{actual infection rate}
  \end{equation}

For a node of state $I$, the infection propagates to its $S$ neighbors with probability $l_I = l(\beta_I)$. For an exposed node who is asymptomatically-infected, the corresponding probability $l_E$ = $l(\beta_E)$. For a node of degree $k$ with $b_E$ neighbors in states $E$ and $b_I$ neighbors in state $I$, the probability of being infected is $q_{b_E,b_I} =1-(1-l_E)^{b_E}(1-l_I)^{b_I}$, where $(1-l_E)^{b_E}(1-l_I)^{b_I}$ is the probability that the node is not infected by any infected neighbors. Thus, the transition probability for a susceptible node being infected at time $t$ is 
 \begin{equation}
 c(t) = \sum_{k=0}^{k_{max}}P(k)\sum_{b_E=0}^{k}\sum_{b_I=0}^{k-b_E}r_{k,b_E,b_I}q_{b_E,b_I},
 \label{StoIprob}
 \end{equation}
 where 
 \begin{equation}
 \begin{aligned}
r_{k,b_E,b_I} = &\binom{k}{b_E}\binom{k-b_E}{b_I}\left\{ [\rho_{E}(t)]^{b_E}[\rho_{I}(t)]^{b_I} \right.\\
 &\left. [1-\rho_{E}(t)-\rho_{I}(t)]^{k-b_E-b_I}\right\}
 \end{aligned}
 \end{equation}
 represents the probability of a node with degree $k$ having $b_E$ neighbors in state $E$ and $b_I$ neighbors in state $I$ at time $t$. 
 
 Then, we can obtain the following equations to describe the dynamics on the \textit{Disease Transmission Layer} and the \textit{Information Diffusion Layer} in Fig.~\ref{fig:3_layer_model} 

\begin{equation}
\begin{aligned} 
\rho_{S}(t+1)-\rho_{S}(t) &= - c(t)\rho_{S}(t),\\
\rho_{E}(t+1)-\rho_{E}(t) &= -\sigma\rho_{E}(t) + c(t)\rho_{S}(t),\\
\rho_{I}(t+1)-\rho_{I}(t) &= -\gamma\rho_{I}(t) + \sigma\rho_{E}(t),\\
\rho_{R}(t+1)-\rho_{R}(t) &= \gamma\rho_{I}(t),\\
\rho_{IG}(t+1)-\rho_{IG}(t) &= - h(t)\rho_{IG}(t),\\
\rho_{SP}(t+1)-\rho_{SP}(t) &= h(t)\rho_{IG}(t)[a\alpha_o+(1-a)\alpha_u],\\
\rho_{SF}(t+1)-\rho_{SF}(t) &= h(t)\rho_{IG}(t)[a(1-\alpha_o)+(1-a)(1-\alpha_u)].
\end{aligned}
\label{diff_equ}
\end{equation}

The dynamics on the \textit{Behavior Change Layer} are described by Eq.~(\ref{behavior change}) and Eq.~(\ref{N}). Information diffusion influences the behavior change by determining the fraction of individuals who are being aware of the disease and adopting self-protection accordingly [From Eq.~(\ref{behavior change})]. Disease transmission is subsequently affected by changing the actual infection rate based on the outcome of behavior change  [From Eq.~(\ref{actual infection rate})].

\section{Results}\label{Results}
We perform extensive Monte Carlo simulations to validate the mean-field approximation results obtained by Eq.~(\ref{diff_equ}). For simplicity, we assume that $\alpha_o = p_o = 0.99$ and $\alpha_u = p_u = 0.01$, indicating an extremely high probability to inform others and to change behavior for ``over-reacting" nodes and a much lower probability for ``under-reacting" nodes. 

\begin{figure}[ht]
\includegraphics[width=1.0\linewidth]{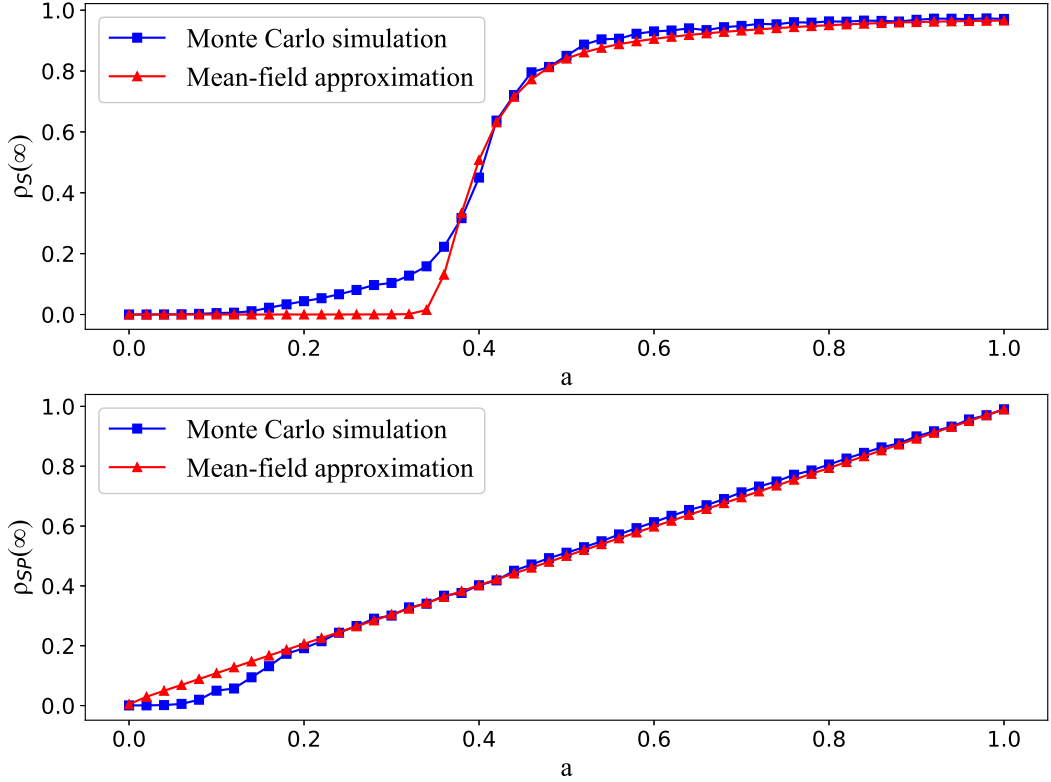}
\caption{\label{fig:SimuAndAna}Comparison of the Monte Carlo simulation and the mean-field approximation results of the relationship between the fraction of susceptible nodes at the stationary state $\rho_{S}(\infty)$ and the fraction of ``over-reacting" nodes $a$ (top panel), and the relationship between the fraction of $Spreaders$ at the stationary state $\rho_{SP}(\infty)$ and $a$ (bottom panel). $\beta_I=0.5, \beta_E=0, \sigma=0.8,\gamma=0.5$, $\eta=0.1$, $\alpha_o=p_o=0.99$, $\alpha_u=p_u=0.01$.}
\end{figure}

We present the comparison between the Monte Carlo simulation and the mean-field approximation results in Fig.~\ref{fig:SimuAndAna} for a three-layer network of 2,000 nodes with a mean degree of 15. We adopt the synchronous updating method \cite{fennell2016limitations} in the Monte Carlo simulations to mimic the discrete-time transmission processes in our model. All nodes are selected to update their states simultaneously in each time step. The time step is set as 1. At the beginning of each simulation, we randomly select a node to be the initially infected node and another to be the first $Spreader$. The updating process stops when there is no infected node in the network. Each point in Fig.~\ref{fig:SimuAndAna} is obtained by averaging 50 Monte Carlo simulations. As illustrated in Fig.~\ref{fig:SimuAndAna}, we observe a high $R^2$ of 0.990 for the top panel (the fraction of susceptible nodes at the stationary state $\rho_{S}(\infty)$), and 0.994 for the bottom panel (the fraction of $Spreaders$ at the stationary state  $\rho_{SP}(\infty)$). This finding validates that the equations in Eq.~(\ref{diff_equ}) can effectively capture the dynamics of the model. Here, the stationary state refers to the state at $t\rightarrow\infty$. Moreover, the fraction of ``over-reacting" nodes $a$ has a linear effect on $\rho_{SP}(\infty)$ but a sigmoid effect on $\rho_{S}(\infty)$.

We examine the degree to which the ``over-reacting" nodes influence the disease dynamics under different situations. Analytical solutions for Eq.~(\ref{diff_equ}) are difficult to obtain. Thus, full phase diagrams are used to illustrate the results. We adopt the same three-layer network mentioned in the aforementioned Monte Carlo simulations.

\begin{figure*}[htbp]
\centering
\includegraphics[width=1.0\linewidth]{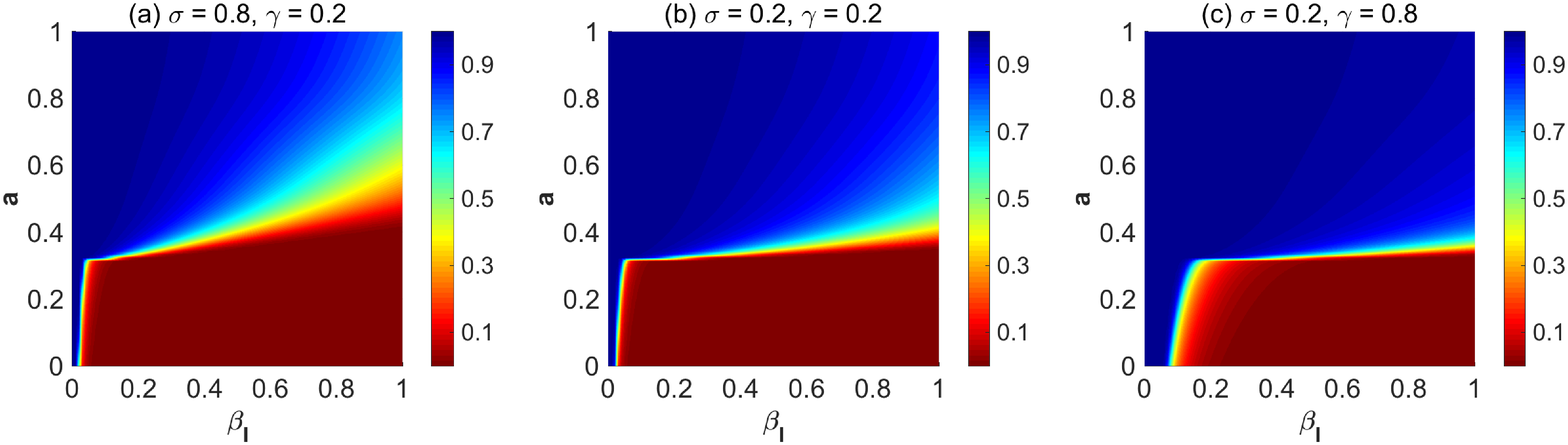}
\caption{Full phase diagrams $a-\beta_I$ of the fraction of susceptible nodes at the stationary state $\rho_{S}(\infty)$ for different values of $\sigma$ and $\gamma$ without asymptomatically-infected cases, where $\eta=0.1$, $\alpha_o=p_o=0.99$, $\alpha_u=p_u=0.01$.}
\label{fig:no asymptomatic infection}
\end{figure*}

First, we focus on the diseases without asymptomatically-infected cases (i.e., $\beta_E=0$). When $\beta_E=0$, the \textit{Diseases Transmission Layer} is essentially an SEIR model and has similar patterns with the standard models given fixed values of $a$ \cite{allen2000comparison,hethcote2000mathematics}. The fraction of susceptible nodes at the stationary state $\rho_{S}(\infty)$ is shown in Fig.~\ref{fig:no asymptomatic infection} concerning the infection rate $\beta_I$ and the fraction of ``over-reacting" nodes $a$. Similar to Fig.~\ref{fig:SimuAndAna}, $a$ has a sigmoid effect on $\rho_{S}(\infty)$ as shown in Fig.~\ref{fig:no asymptomatic infection}. We find that $\rho_{S}(\infty)$ generally has three phases: When $a$ is small, $\rho_{S}(\infty)$ is close to 0, and when $a$ is large, $\rho_{S}(\infty)$ is close to 1. A threshold triggers the rapid increase in $\rho_{S}(\infty)$. The exact value of the threshold cannot be analytically obtained. This indicates that by increasing the fraction of ``over-reacting" nodes through highlighting the transmissibility and severity of the disease, we can largely prevent the disease outbreak. As long as the fraction of ``over-reacting" nodes reaches the threshold, further highlight this fraction is not necessary. This finding shows that the disease outbreak is preventable when the media and opinion leaders are playing effective roles of the whistle-blower. 
 
\begin{figure*}[htbp]
\centering
\includegraphics[width=1.0\linewidth]{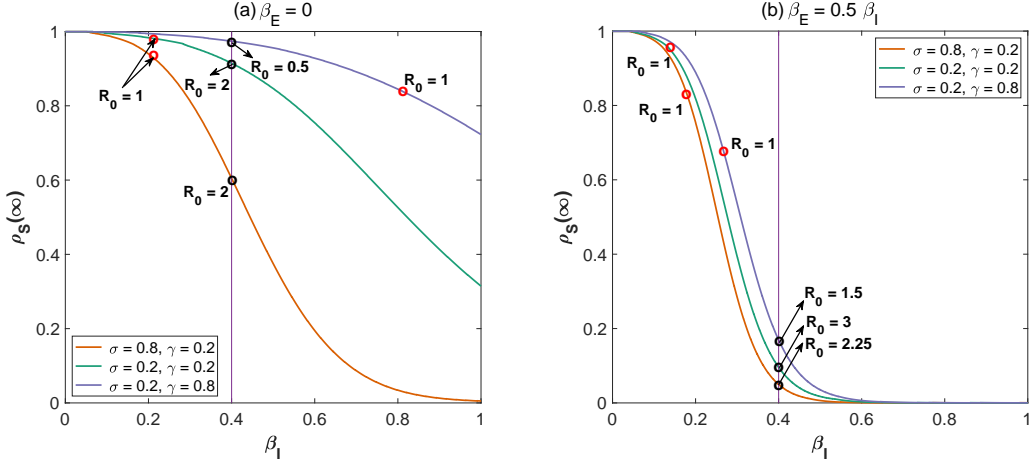}
\caption{Fraction of susceptible nodes at the stationary state $\rho_{S}(\infty)$ with respect to $\beta_I$ for different values of $\sigma$ and $\gamma$, where $a=0.4$, $\eta=0.1$, $\alpha_o=p_o=0.99$, $\alpha_u=p_u=0.01$.}
\label{fig:nobetaE_and_betaE05}
\end{figure*}

More specifically, when $\beta_I$ is extremely small, all possible values of $a$ yield to the fully prevented scenario (blue). As $\beta_I$ increases, the range of $a$ rapidly increases and yields the full outbreak scenario (red). When $\beta_I$ keeps increasing, the range of $a$ yielding to full outbreak scenario gradually increases. We define the critical value of $\beta_I$ as the value leading to the basic reproduction number $R_0=\beta_I/\gamma=1$, below which the epidemic is under control. As shown in panel (a) of Fig.~\ref{fig:nobetaE_and_betaE05}, where $a$ is fixed, the higher the recovery rate $\gamma$, the larger the critical value of $\beta_I$. Given the same $\beta_I$, we can observe an increase in $\rho_{S}(\infty)$ with a longer incubation period $1/\sigma$ or a smaller $R_0$ (larger $\gamma$), see panel (a) of Fig.~\ref{fig:nobetaE_and_betaE05}. It might
be because a longer incubation period can provide more time for the awareness to be transmitted among people, and a smaller $R_0$ means a less transmissible disease. Thus, a smaller fraction of ``over-reacting" nodes is needed to prevent the full outbreak.

\begin{figure*}[htbp]
\centering
\includegraphics[width=1.0\linewidth]{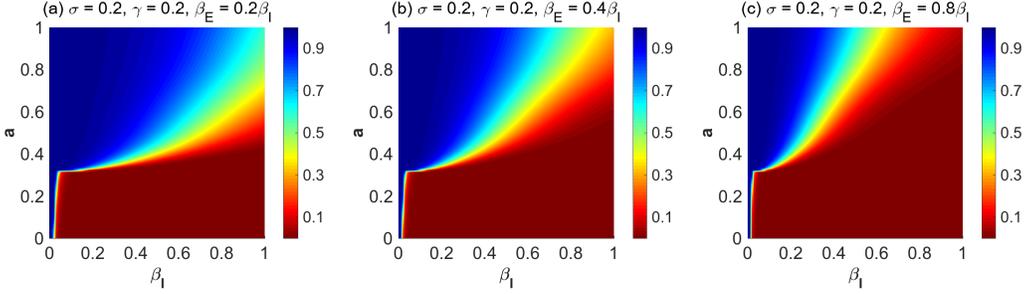}
\caption{Full phase diagrams $a-\beta_I$ of the fraction of susceptible nodes at the stationary state $\rho_{S}(\infty)$ for $\beta_E=0.2\beta_I$, $\beta_E=0.4\beta_I$, $\beta_E=0.8\beta_I$, where $\eta=0.1$, $\alpha_o=p_o=0.99$, $\alpha_u=p_u=0.01, \sigma=0.2,\gamma=0.2$.}
\label{fig:asymptomatic infection}
\end{figure*}

Second, we consider the diseases with asymptomatically-infected cases. Given that not all exposed individuals are asymptomatically-infected, and the asymptomatically-infected cases are often not more infectious than symptomatically-infected cases \cite{wilder2005asymptomatic,NEJMc2001468,papenburg2010household,leeclinical}, we denote the asymptomatic infection rate as $\beta_E=\mu\beta_I$, where $\mu\in(0,1)$. We report the fraction of susceptible nodes at the stationary state $\rho_{S}(\infty)$ in Fig.~\ref{fig:asymptomatic infection} where $\mu \in \{0.2, 0.4, 0.8\}$. The values of other parameters are the same as those in panel (b) of Fig.~\ref{fig:no asymptomatic infection}, where the asymptomatic infection is not considered. We find that when asymptomatic infections exist, the range of $a$ yielding to the full outbreak scenario is larger, and the effect becomes more significant with a larger value of $\mu$. In the extreme case when $\mu\rightarrow1$, the \textit{Disease Transmission Layer} essentially becomes an SIR model with longer infectious period than the original SEIR model. These results indicate that asymptomatic infections make it harder, sometimes even impossible, to control the epidemic by only increasing the value of $a$ (panel (c) of Fig.~\ref{fig:asymptomatic infection}). 

\begin{figure*}[htbp]
\centering
\includegraphics[width=1.0\linewidth]{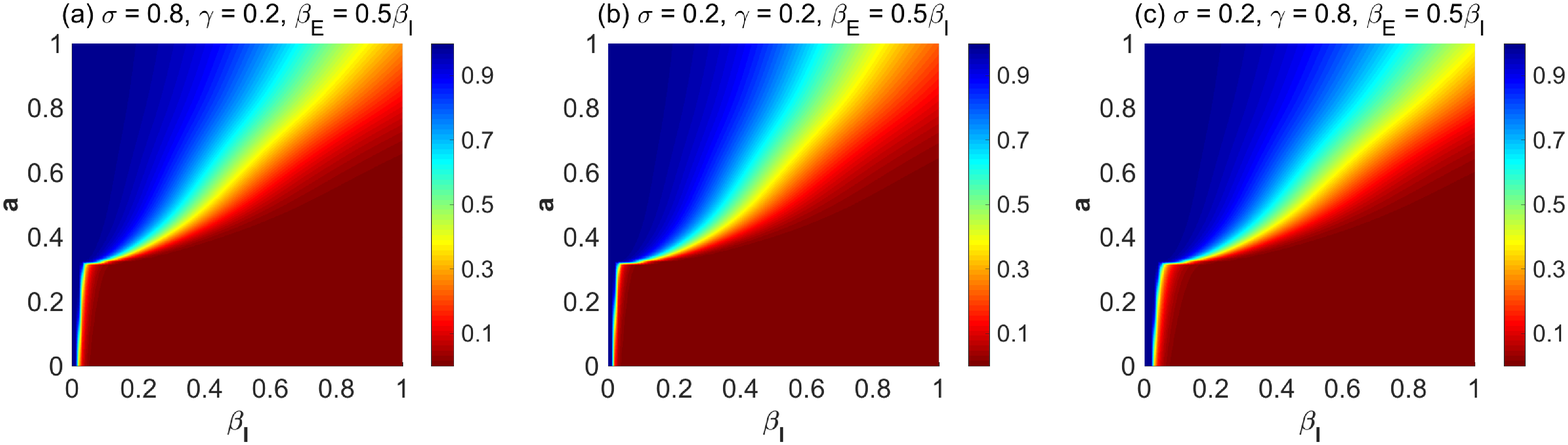}
\caption{Full phase diagrams $a-\beta_I$ of the fraction of susceptible nodes at the stationary state $\rho_{S}(\infty)$ for different values of $\sigma$ and $\gamma$ with asymptomatically-infected cases, where $\beta_E=0.5\beta_I$, $\eta=0.1$, $\alpha_o=p_o=0.99$, $\alpha_u=p_u=0.01$.}
\label{fig:asymptomatic infection_sigma_gamma}
\end{figure*}

Furthermore, we explore the effects of ``over-reacting" nodes with various parameter settings on the \textit{Disease Transmission Layer} (representing different diseases) in Fig.~\ref{fig:asymptomatic infection_sigma_gamma}. Compared with the cases where $\beta_E = 0$, the existence of asymptomatic infections reduces the critical values of $\beta_I$ and the values of $\rho_S(\infty)$, indicating that epidemic control is more difficult (see Fig.~\ref{fig:nobetaE_and_betaE05}). With the existence of asymptomatically-infected cases, $R_0$ = $\beta_I / \gamma$ + $\beta_E / \sigma$ \cite{oliveira2020refined}. We find that not only the higher recovery rate $\gamma$ but also the shorter incubation period $1/\sigma$ can increase the critical value of $\beta_I$. A smaller $R_0$ does not always lead to a larger $\rho_{S}(\infty)$ in such cases. As illustrated in panel (b) of Fig.~\ref{fig:nobetaE_and_betaE05}, given the same $\beta_I$, when $\gamma$ increases to 0.8 from 0.2, $R_0$ drops to 1.5 from 3. Thus, $\rho_{S}(\infty)$ increases for a less transmissible disease. However, when 
$\sigma$ increases to 0.8 (shorter incubation period), $R_0$ decreases to 2.25 from 3 (less transmissible). Counterintuitively, $\rho_{S}(\infty)$ decreases to 0.05 from 0.096. That is because a larger $\sigma$ not only leads to a smaller $R_0$ but also indicates a shorter incubation time for the awareness to spread out; therefore, fewer people are aware and protected.

\begin{figure*}[htbp]
\centering
\includegraphics[width=1.0\linewidth]{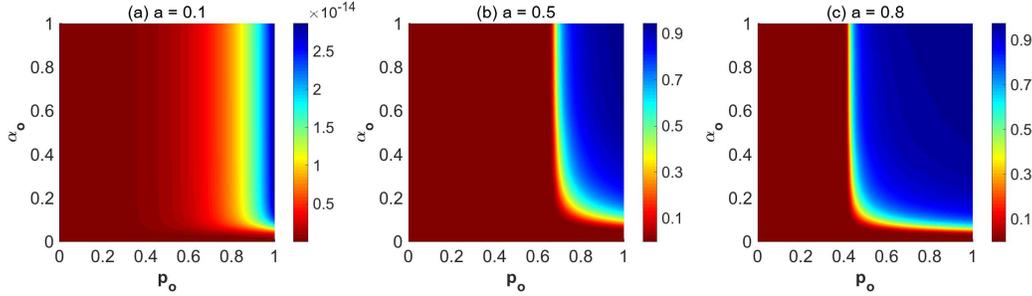}
\caption{Full phase diagrams $\alpha_o-p_o$ of the fraction of susceptible nodes at the stationary state $\rho_{S}(\infty)$ for different values of $a$, where $\beta_I=0.5, \beta_E=0, \sigma=\gamma=0.2, \eta=0.1$.}
\label{fig:p_and_alpha}
\end{figure*}

Third, we further clarify the effect of ``over-reacting" nodes on epidemic control for different values of $\alpha$ and $p$ in Fig.~\ref{fig:p_and_alpha}. For simplicity, we assume $\alpha_u = \frac{1}{2} \alpha_o$ and $p_u=\frac{1}{2} p_o$.  Fewer nodes will be infected when $\alpha$ and $p$ increase. When the value of $a$ increases, the epidemic is easier to control (a larger blue region in the figure). A small transitioning space occurs between controlled (blue) and outbreak (red) scenarios, indicating that the epidemic can either be well contained or will likely to infect the majority of people in the population. 

\begin{figure}[htbp]
\includegraphics[width=1.0\linewidth]{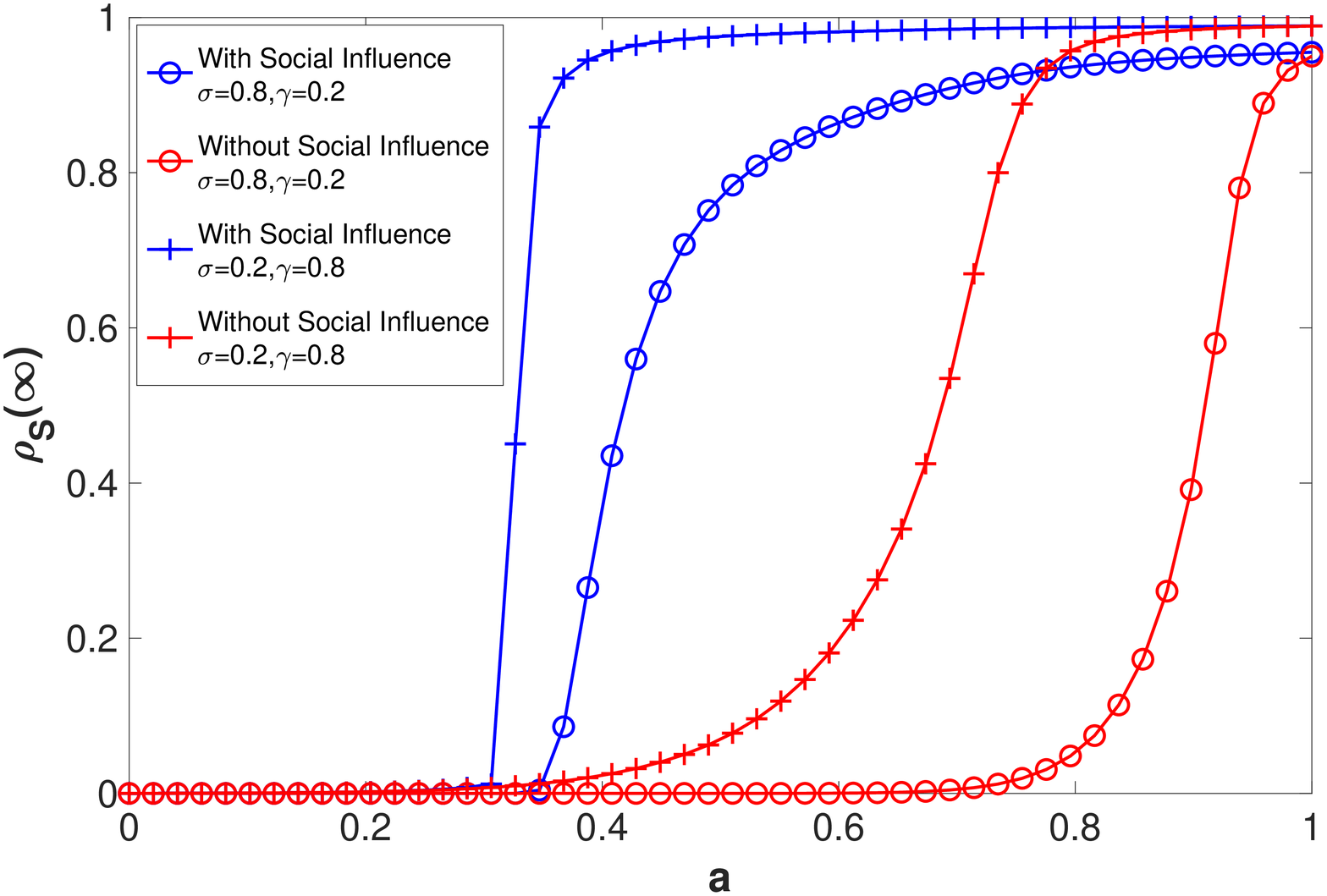}
\caption{\label{fig:social_influence} Comparison of the fraction of susceptible nodes at the stationary state $\rho_{S}(\infty)$ with respect to $a$ for two diseases, where $\eta=0.1$, $\alpha_o=p_o=0.99$, $\alpha_u=p_u=0.01$, $\beta_I=0.5$, $\beta_E=0$.}
\end{figure}

Finally, we examine the fraction of susceptible nodes at the stationary state $\rho_{S}(\infty)$ in two scenarios: with and without social influence on behavior change. Here, the scenario without social influence is modeled by setting $W=0$ regardless of the behavior of the node's neighbors. We consider two diseases with different parameter settings in Fig. \ref{fig:social_influence}. With the social influence on behavior change, we can achieve the fully controlled result ($\rho_{S}(\infty)\rightarrow1$) with a lower value of $a$ for both diseases, indicating that we can effectively utilize the social influence among people to enhance the disease prevention.

\begin{figure*}[htbp]
\centering
\includegraphics[width=1.0\linewidth]{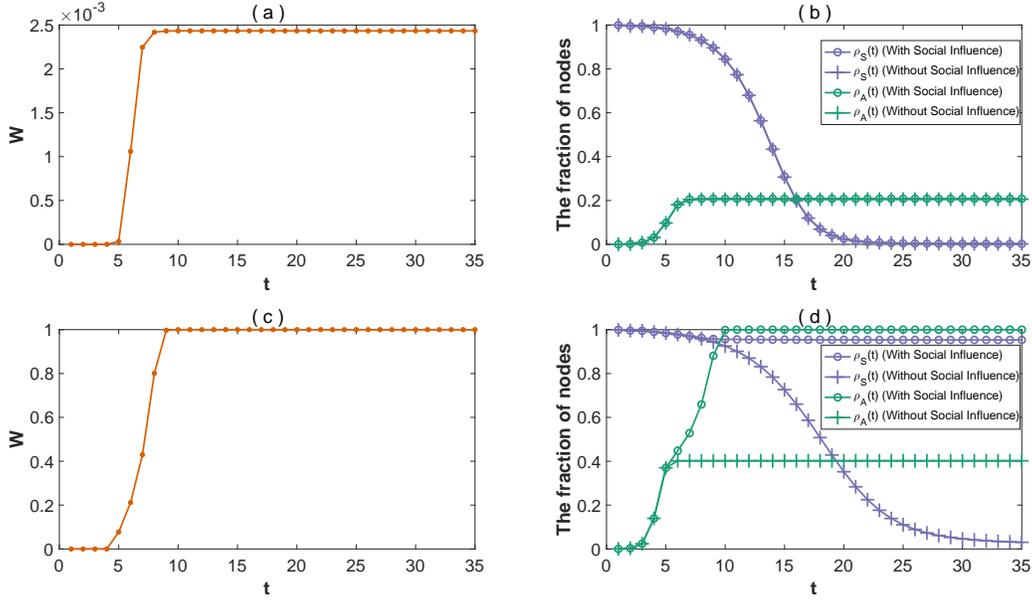}
\caption{Effect of social influence on the epidemic with different fractions of ``over-reacting" nodes. Panel (a) and panel (c): probability of changing behavior due to the social influence from neighbors ($W$) at time $t$. Panel (b) and panel (d): fraction of susceptible nodes ($\rho_S(t)$) and fraction of nodes having adopted self-protection ($\rho_A(t)$) over time. $a=0.2$ in panel (a) and (b), $a=0.4$ in panel (c) and (d). $\beta_I=0.5$, $\beta_E=0$, $\sigma=0.2$, $\gamma=0.8$, $\eta=0.1$, $\alpha_o=p_o=0.99$, $\alpha_u=p_u=0.01$.}
\label{fig:explain}
\end{figure*}

To further clarify such a significant effect on $\rho_{S}(\infty)$, we focus on one specific disease setting mentioned in Fig.~\ref{fig:social_influence} and analyze the fraction of nodes in state $S$, $A$ ($\rho_{S}(t)$, $\rho_{A}(t)$) and the probability of changing behavior due to the social influence from neighbors ($W$) over time in Fig.~\ref{fig:explain}. Here, we consider two scenarios, where $a=0.2$ and $a=0.4$, respectively. We observe that $W$ increases rapidly over time and reaches an equilibrium state for both cases. Larger $a$ leads to a higher value of $W$ at the equilibrium state. Assuming that the \textit{Information Diffusion Layer} reaches the equilibrium state at time $t_I$, $W$ reaches the equilibrium state at time $t_W$, set $T=max\{t_I, t_W\}$, we obtain 
 \begin{equation}
 \begin{aligned}
 \rho_A(T) =& [\rho_{SP}(T)+\rho_{SF}(T)]\left.\{ap_o +(1-a)p_u+\right.\\
 &\left.[a(1-p_o)+(1-a)(1-p_u)]W\right.\}\\
 =& [\rho_{SP}(T)+\rho_{SF}(T)][ap_o +(1-a)p_u] + [\rho_{SP}(T)\\ 
 &+\rho_{SF}(T)][a(1-p_o)+(1-a)(1-p_u)]W
 \end{aligned}
 \label{behavior change infity}
 \end{equation}
from Eq.~(\ref{behavior change}), indicating that the fraction of people who have adopted self-protection, $\rho_A(T)$, can be increased with the help of social influence. In addition,
\begin{equation}
\rho_{IG}(T) = \rho_{IG}(T)[1-h(T)],
\end{equation}
which indicates that $\rho_{IG}(T)=0$ or $h(T)=0$. $h(T)=0$ only if $\rho_{SP}(T)=0$ according to Eq.~(\ref{ht}). We can derive that if $\overline{k}[a\alpha_o+(1-a)\alpha_u]>1$, where $\overline{k}$ is the mean degree of the network, then information can always spread out ($\rho_{SP}(T)>0$) \cite{ruan2018information}. In such cases, $h(T)\neq0$, thus, $\rho_{IG}(T)$ $=0$, indicating that $\rho_{SP}(T)+\rho_{SF}(T)=1$. Therefore, Eq.~(\ref{behavior change infity}) can be simplified as follows 
\begin{equation}
\rho_A(T)=ap_o+(1-a)p_u+[a(1-p_o)+(1-a)(1-p_u)]W,
\end{equation}
if $\overline{k}[a\alpha_o+(1-a)\alpha_u]>1$. 

When $a=0.2$, the value of $W$ at $t=T$ is small, indicating that social influence has little effect on the behavior change. In such cases, $\overline{k}[a\alpha_o+(1-a)\alpha_u]=3.09>1$, then $\rho_A(T)\approx ap_o$ $+(1-a)p_u=0.206$, which is small, indicating that the epidemic cannot be well contained. However, when $a=0.4$, $W$ reaches $1$ rapidly, and $\overline{k}[a\alpha_o+(1-a)\alpha_u]=6.03>1$, then $\rho_A(T)=1$. Thus, the actual infection rate $l(\beta)=\beta\eta^2$ is close to $0$ for small $\eta$ according to Eq.~(\ref{actual infection rate}). As a result, the transition probability for a susceptible node being infected is $c(T)\rightarrow0$ according to Eq.~(\ref{StoIprob}), indicating that the epidemic is approaching the end. Since $\rho_{S}(t)$ is decreasing over time, the earlier $c(t)$ becomes $0$, the fewer people are infected. Therefore, the fraction of susceptible nodes tends to remain high since $t=T$ with the existence of social influence, which happens earlier than that without social influence (panel (d) in Fig.~\ref{fig:explain}). These results indicate that social influence on behavior change can significantly accelerate behavior change and lead to a smaller epidemic outbreak.

\section{Conclusion}\label{Conclusion}
In this study, we develop a \textbf{h}eterogeneous \textbf{D}isease-\textbf{B}ehavior-\textbf{I}nformation (\textbf{hDBI}) transmission model to characterize the heterogeneous processes of information diffusion, behavior change, and disease transmission on social networks. We adopt the mean-field approximation approach to obtain analytical results and perform extensive Monte Carlo simulations to examine the patterns of disease transmission in the presence of information diffusion and behavior change among people. We find that (a) disease awareness plays a central role in preventing the disease outbreak; (b) a reasonable fraction of ``over-reacting" nodes are needed to effectively control the epidemic; (c) a smaller $R_0$ always leads to a smaller epidemic outbreak without symptomatically-infected cases. This scenario would have different effects with asymptomatically-infected cases because a smaller $R_0$ might result from a shorter incubation time. As a result, people have a shorter period to become aware and adopt self-protection; (d) social influence on behavior change can significantly decrease the outbreak size. 

In practice, with the absence of vaccines and stringent control measures, the epidemic can still be well contained when people are aware of the disease and adopt proper self-protection. The media and opinion leaders play a key role in people's disease awareness. If transmissibility and severity are understated by them, then more people will remain ``under-reacting" and thus will be infected eventually. However, if they state unreasonably high transmissibility and severity of the disease, the ``crying wolf" effect could result in people losing confidence in the public health system. Further research is needed to identify the optimal degree of transmissibility and severity stated by the media and opinion leaders. This research has limitations. In the current model, we assume an unchanged topological structure of the multiplex network, but uninfected nodes might break connections with infected nodes and form new connections with other uninfected nodes to avoid being infected during epidemics. Incorporating the adaptive network schemes into the current model by introducing a rewiring probability for uninfected nodes will be our future work. In the COVID-19 context, many countries have seen protests against the lockdown measure, indicating that $Stiflers$ could even become another type of competing $Spreaders$ that influence $Ignorants$ to become $Stiflers$. Modeling the competing behaviors during epidemics is an interesting topic for further exploration.

\bibliographystyle{unsrt}  
\bibliography{ref} 

\begin{thebibliography}{10}

\bibitem{Senn:2009}
WHO.
\newblock {\em Protect yourself and others}, 2020.
\newblock
  \url{http://www.emro.who.int/health-topics/corona-virus/protect-yourself-and-others.html}.

\bibitem{ferguson2007capturing}
Neil Ferguson.
\newblock Capturing human behaviour.
\newblock {\em Nature}, 446(7137):733--733, 2007.

\bibitem{Rubinb2651}
G~James Rubin, Richard Aml{\^o}t, Lisa Page, and Simon Wessely.
\newblock Public perceptions, anxiety, and behaviour change in relation to the
  swine flu outbreak: cross sectional telephone survey.
\newblock {\em BMJ}, 339, 2009.

\bibitem{Lau864}
J~T~F Lau, X~Yang, H~Tsui, and J~H Kim.
\newblock Monitoring community responses to the sars epidemic in hong kong:
  from day 10 to day 62.
\newblock {\em Journal of Epidemiology \& Community Health}, 57(11):864--870,
  2003.

\bibitem{RePEc:uwp:jhriss:v:31:y:1996:i:3:p:611-630}
Tomas Philipson.
\newblock {Private Vaccination and Public Health: An Empirical Examination for
  U.S. Measles}.
\newblock {\em Journal of Human Resources}, 31(3):611--630, 1996.

\bibitem{article_AIDS}
Kate Macintyre, Lisanne Brown, and Stephen Sosler.
\newblock It's not what you know, but who you knew: Examining the relationship
  between behavior change and aids mortality in africa.
\newblock {\em AIDS education and prevention : official publication of the
  International Society for AIDS Education}, 13:160--74, 05 2001.

\bibitem{collinson2014modelling}
Shannon Collinson and Jane~M Heffernan.
\newblock Modelling the effects of media during an influenza epidemic.
\newblock {\em BMC public health}, 14(1):376, 2014.

\bibitem{van2020using}
Jay~J Van~Bavel, Katherine Baicker, Paulo~S Boggio, Valerio Capraro, Aleksandra
  Cichocka, Mina Cikara, Molly~J Crockett, Alia~J Crum, Karen~M Douglas,
  James~N Druckman, et~al.
\newblock Using social and behavioural science to support covid-19 pandemic
  response.
\newblock {\em Nature Human Behaviour}, pages 1--12, 2020.

\bibitem{9043580}
L.~{Li}, Q.~{Zhang}, X.~{Wang}, J.~{Zhang}, T.~{Wang}, T.~{Gao}, W.~{Duan},
  K.~K. {Tsoi}, and F.~{Wang}.
\newblock Characterizing the propagation of situational information in social
  media during covid-19 epidemic: A case study on weibo.
\newblock {\em IEEE Transactions on Computational Social Systems},
  7(2):556--562, 2020.

\bibitem{perra2012activity}
Nicola Perra, Bruno Gon{\c{c}}alves, Romualdo Pastor-Satorras, and Alessandro
  Vespignani.
\newblock Activity driven modeling of time varying networks.
\newblock {\em Scientific reports}, 2:469, 2012.

\bibitem{wang2014multiple}
Zhigang Wang, Haifeng Zhang, and Zhen Wang.
\newblock Multiple effects of self-protection on the spreading of epidemics.
\newblock {\em Chaos, Solitons \& Fractals}, 61:1--7, 2014.

\bibitem{ruan2003dynamical}
Shigui Ruan and Wendi Wang.
\newblock Dynamical behavior of an epidemic model with a nonlinear incidence
  rate.
\newblock {\em Journal of Differential Equations}, 188(1):135--163, 2003.

\bibitem{WANG20151}
Zhen Wang, Michael~A. Andrews, Zhi-Xi Wu, Lin Wang, and Chris~T. Bauch.
\newblock Coupled disease–behavior dynamics on complex networks: A review.
\newblock {\em Physics of Life Reviews}, 15:1 -- 29, 2015.

\bibitem{kabir2020impact}
KM~Ariful Kabir, Kazuki Kuga, and Jun Tanimoto.
\newblock The impact of information spreading on epidemic vaccination game
  dynamics in a heterogeneous complex network-a theoretical approach.
\newblock {\em Chaos, Solitons \& Fractals}, 132:109548, 2020.

\bibitem{han2020individuals}
Dun Han, Qi~Shao, Dandan Li, and Mei Sun.
\newblock How the individuals’ risk aversion affect the epidemic spreading.
\newblock {\em Applied Mathematics and Computation}, 369:124894, 2020.

\bibitem{guo2013epidemic}
Dongchao Guo, Stojan Trajanovski, Ruud van~de Bovenkamp, Huijuan Wang, and Piet
  Van~Mieghem.
\newblock Epidemic threshold and topological structure of
  susceptible-infectious-susceptible epidemics in adaptive networks.
\newblock {\em Physical Review E}, 88(4):042802, 2013.

\bibitem{fenichel2011adaptive}
Eli~P Fenichel, Carlos Castillo-Chavez, M~Graziano Ceddia, Gerardo Chowell,
  Paula A~Gonzalez Parra, Graham~J Hickling, Garth Holloway, Richard Horan,
  Benjamin Morin, Charles Perrings, et~al.
\newblock Adaptive human behavior in epidemiological models.
\newblock {\em Proceedings of the National Academy of Sciences},
  108(15):6306--6311, 2011.

\bibitem{article_Media_Reports}
Shannon Collinson, Kamran Khan, and Jane Heffernan.
\newblock The effects of media reports on disease spread and important public
  health measurements.
\newblock {\em PloS one}, 10:e0141423, 11 2015.

\bibitem{funk2009spread}
Sebastian Funk, Erez Gilad, Chris Watkins, and Vincent~AA Jansen.
\newblock The spread of awareness and its impact on epidemic outbreaks.
\newblock {\em Proceedings of the National Academy of Sciences},
  106(16):6872--6877, 2009.

\bibitem{PhysRevLett.96.208701}
Thilo Gross, Carlos J.~Dommar D'Lima, and Bernd Blasius.
\newblock Epidemic dynamics on an adaptive network.
\newblock {\em Phys. Rev. Lett.}, 96:208701, May 2006.

\bibitem{article}
Qingchu Wu, Xinchu Fu, Michael Small, and Xin-Jian Xu.
\newblock The impact of awareness on epidemic spreading in networks.
\newblock {\em Chaos (Woodbury, N.Y.)}, 22:013101, 03 2012.

\bibitem{CHEN2009125}
Frederick~H. Chen.
\newblock Modeling the effect of information quality on risk behavior change
  and the transmission of infectious diseases.
\newblock {\em Mathematical Biosciences}, 217(2):125 -- 133, 2009.

\bibitem{article_Layered_Complex_Networks}
Maciej Kurant and Patrick Thiran.
\newblock Layered complex networks.
\newblock {\em Physical review letters}, 96:138701, 05 2006.

\bibitem{Multilayer}
Mikko Kivelä, Alex Arenas, Marc Barthelemy, James~P. Gleeson, Yamir Moreno,
  and Mason~A. Porter.
\newblock {Multilayer networks}.
\newblock {\em Journal of Complex Networks}, 2(3):203--271, 07 2014.

\bibitem{PhysRevLett.111.128701}
Clara Granell, Sergio G\'omez, and Alex Arenas.
\newblock Dynamical interplay between awareness and epidemic spreading in
  multiplex networks.
\newblock {\em Phys. Rev. Lett.}, 111:128701, Sep 2013.

\bibitem{PAN201845}
Yaohui Pan and Zhijun Yan.
\newblock The impact of multiple information on coupled awareness-epidemic
  dynamics in multiplex networks.
\newblock {\em Physica A: Statistical Mechanics and its Applications}, 491:45
  -- 54, 2018.

\bibitem{PhysRevE.91.012822}
Quantong Guo, Xin Jiang, Yanjun Lei, Meng Li, Yifang Ma, and Zhiming Zheng.
\newblock Two-stage effects of awareness cascade on epidemic spreading in
  multiplex networks.
\newblock {\em Phys. Rev. E}, 91:012822, Jan 2015.

\bibitem{KABIR2019118}
K.M.~Ariful Kabir, Kazuki Kuga, and Jun Tanimoto.
\newblock Analysis of sir epidemic model with information spreading of
  awareness.
\newblock {\em Chaos, Solitons and Fractals}, 119:118 -- 125, 2019.

\bibitem{mao2014modeling}
Liang Mao.
\newblock Modeling triple-diffusions of infectious diseases, information, and
  preventive behaviors through a metropolitan social network—an agent-based
  simulation.
\newblock {\em Applied Geography}, 50:31--39, 2014.

\bibitem{granell2014competing}
Clara Granell, Sergio G{\'o}mez, and Alex Arenas.
\newblock Competing spreading processes on multiplex networks: awareness and
  epidemics.
\newblock {\em Physical review E}, 90(1):012808, 2014.

\bibitem{perra2011towards}
Nicola Perra, Duygu Balcan, Bruno Gon{\c{c}}alves, and Alessandro Vespignani.
\newblock Towards a characterization of behavior-disease models.
\newblock {\em PloS one}, 6(8), 2011.

\bibitem{liu2007media}
Rongsong Liu, Jianhong Wu, and Huaiping Zhu.
\newblock Media/psychological impact on multiple outbreaks of emerging
  infectious diseases.
\newblock {\em Computational and Mathematical Methods in Medicine},
  8(3):153--164, 2007.

\bibitem{wang2014filippov}
Aili Wang and Yanni Xiao.
\newblock A filippov system describing media effects on the spread of
  infectious diseases.
\newblock {\em Nonlinear Analysis: Hybrid Systems}, 11:84--97, 2014.

\bibitem{dubey2015proceedings}
Preeti Dubey, Balram Dubey, and Uma Dubey.
\newblock Proceedings of the international symposium on mathematical and
  computational biology.
\newblock {\em BIOMAT 2015}, 2015.

\bibitem{CRAMER2016739}
Emily~M. Cramer, Hayeon Song, and Adam~M. Drent.
\newblock Social comparison on facebook: Motivation, affective consequences,
  self-esteem, and facebook fatigue.
\newblock {\em Computers in Human Behavior}, 64:739 -- 746, 2016.

\bibitem{song2018global}
Pengfei Song and Yanni Xiao.
\newblock Global hopf bifurcation of a delayed equation describing the lag
  effect of media impact on the spread of infectious disease.
\newblock {\em Journal of mathematical biology}, 76(5):1249--1267, 2018.

\bibitem{song2019analysis}
Pengfei Song and Yanni Xiao.
\newblock Analysis of an epidemic system with two response delays in media
  impact function.
\newblock {\em Bulletin of mathematical biology}, 81(5):1582--1612, 2019.

\bibitem{sjoberg2000factors}
Lennart Sj{\^o}berg.
\newblock Factors in risk perception.
\newblock {\em Risk analysis}, 20(1):1--12, 2000.

\bibitem{wahlberg2000risk}
Anders~AF Wahlberg and Lennart Sjoberg.
\newblock Risk perception and the media.
\newblock {\em Journal of risk research}, 3(1):31--50, 2000.

\bibitem{RENNER2015702}
Ralf~Schmälzle Britta~Renner, Martina~Gamp and Harald~T. Schupp.
\newblock Health risk perception.
\newblock In James~D. Wright, editor, {\em International Encyclopedia of the
  Social and Behavioral Sciences (Second Edition)}, pages 702 -- 709. Elsevier,
  Oxford, 2015.

\bibitem{ruan2018information}
Zhongyuan Ruan, Jinbao Wang, Qi~Xuan, Chenbo Fu, and Guanrong Chen.
\newblock Information filtering by smart nodes in random networks.
\newblock {\em Physical Review E}, 98(2):022308, 2018.

\bibitem{ruan2012epidemic}
Zhongyuan Ruan, Ming Tang, and Zonghua Liu.
\newblock Epidemic spreading with information-driven vaccination.
\newblock {\em Physical Review E}, 86(3):036117, 2012.

\bibitem{zhu2020novel}
Na~Zhu, Dingyu Zhang, Wenling Wang, Xingwang Li, Bo~Yang, Jingdong Song, Xiang
  Zhao, Baoying Huang, Weifeng Shi, Roujian Lu, et~al.
\newblock A novel coronavirus from patients with pneumonia in china, 2019.
\newblock {\em New England Journal of Medicine}, 2020.

\bibitem{wang2020novel}
Chen Wang, Peter~W Horby, Frederick~G Hayden, and George~F Gao.
\newblock A novel coronavirus outbreak of global health concern.
\newblock {\em The Lancet}, 395(10223):470--473, 2020.

\bibitem{ASMUNDSON2020102196}
Gordon~J.G. Asmundson and Steven Taylor.
\newblock Coronaphobia: Fear and the 2019-ncov outbreak.
\newblock {\em Journal of Anxiety Disorders}, 70:102196, 2020.

\bibitem{kai2020universal}
De~Kai, Guy-Philippe Goldstein, Alexey Morgunov, Vishal Nangalia, and Anna
  Rotkirch.
\newblock Universal masking is urgent in the covid-19 pandemic: Seir and agent
  based models, empirical validation, policy recommendations.
\newblock {\em arXiv preprint arXiv:2004.13553}, 2020.

\bibitem{sahneh2013generalized}
Faryad~Darabi Sahneh, Caterina Scoglio, and Piet Van~Mieghem.
\newblock Generalized epidemic mean-field model for spreading processes over
  multilayer complex networks.
\newblock {\em IEEE/ACM Transactions on Networking}, 21(5):1609--1620, 2013.

\bibitem{wilder2005asymptomatic}
Annelies Wilder-Smith, Monica~D Teleman, Bee~H Heng, Arul Earnest, Ai~E Ling,
  and Yee~S Leo.
\newblock Asymptomatic sars coronavirus infection among healthcare workers,
  singapore.
\newblock {\em Emerging infectious diseases}, 11(7):1142, 2005.

\bibitem{NEJMc2001468}
Camilla Rothe, Mirjam Schunk, Peter Sothmann, Gisela Bretzel, Guenter Froeschl,
  Claudia Wallrauch, Thorbjörn Zimmer, Verena Thiel, Christian Janke, Wolfgang
  Guggemos, Michael Seilmaier, Christian Drosten, Patrick Vollmar, Katrin
  Zwirglmaier, Sabine Zange, Roman Wölfel, and Michael Hoelscher.
\newblock Transmission of 2019-ncov infection from an asymptomatic contact in
  germany.
\newblock {\em New England Journal of Medicine}, 382(10):970--971, 2020.

\bibitem{papenburg2010household}
Jesse Papenburg, Mariana Baz, Marie-{\`E}ve Hamelin, Chantal Rh{\'e}aume, Julie
  Carbonneau, Manale Ouakki, Isabelle Rouleau, Isabelle Hardy, Danuta
  Skowronski, Michel Roger, et~al.
\newblock Household transmission of the 2009 pandemic a/h1n1 influenza virus:
  elevated laboratory-confirmed secondary attack rates and evidence of
  asymptomatic infections.
\newblock {\em Clinical Infectious Diseases}, 51(9):1033--1041, 2010.

\bibitem{hou2020effectiveness}
Can Hou, Jiaxin Chen, Yaqing Zhou, Lei Hua, Jinxia Yuan, Shu He, Yi~Guo, Sheng
  Zhang, Qiaowei Jia, Chenhui Zhao, et~al.
\newblock The effectiveness of quarantine of wuhan city against the corona
  virus disease 2019 (covid-19): A well-mixed seir model analysis.
\newblock {\em Journal of medical virology}, 2020.

\bibitem{wu2020nowcasting}
Joseph~T Wu, Kathy Leung, and Gabriel~M Leung.
\newblock Nowcasting and forecasting the potential domestic and international
  spread of the 2019-ncov outbreak originating in wuhan, china: a modelling
  study.
\newblock {\em The Lancet}, 395(10225):689--697, 2020.

\bibitem{6423227}
F.~{Darabi Sahneh}, C.~{Scoglio}, and P.~{Van Mieghem}.
\newblock Generalized epidemic mean-field model for spreading processes over
  multilayer complex networks.
\newblock {\em IEEE/ACM Transactions on Networking}, 21(5):1609--1620, 2013.

\bibitem{barabasi1999mean}
Albert-L{\'a}szl{\'o} Barab{\'a}si, R{\'e}ka Albert, and Hawoong Jeong.
\newblock Mean-field theory for scale-free random networks.
\newblock {\em Physica A: Statistical Mechanics and its Applications},
  272(1-2):173--187, 1999.

\bibitem{li2012susceptible}
Cong Li, Ruud van~de Bovenkamp, and Piet Van~Mieghem.
\newblock Susceptible-infected-susceptible model: A comparison of n-intertwined
  and heterogeneous mean-field approximations.
\newblock {\em Physical Review E}, 86(2):026116, 2012.

\bibitem{fennell2016limitations}
Peter~G Fennell, Sergey Melnik, and James~P Gleeson.
\newblock Limitations of discrete-time approaches to continuous-time contagion
  dynamics.
\newblock {\em Physical Review E}, 94(5):052125, 2016.

\bibitem{allen2000comparison}
Linda~JS Allen and Amy~M Burgin.
\newblock Comparison of deterministic and stochastic sis and sir models in
  discrete time.
\newblock {\em Mathematical biosciences}, 163(1):1--33, 2000.

\bibitem{hethcote2000mathematics}
Herbert~W Hethcote.
\newblock The mathematics of infectious diseases.
\newblock {\em SIAM review}, 42(4):599--653, 2000.

\bibitem{leeclinical}
Seungjae Lee, Tark Kim, Eunjung Lee, Cheolgu Lee, Hojung Kim, Heejeong Rhee,
  Se~Yoon Park, Hyo-Ju Son, Shinae Yu, Jung~Wan Park, et~al.
\newblock Clinical course and molecular viral shedding among asymptomatic and
  symptomatic patients with sars-cov-2 infection in a community treatment
  center in the republic of korea.
\newblock {\em JAMA Internal Medicine}.

\bibitem{oliveira2020refined}
Goncalo Oliveira.
\newblock Refined compartmental models, asymptomatic carriers and covid-19.
\newblock {\em arXiv preprint arXiv:2004.14780}, 2020.

\end{thebibliography}

\end{document}